\documentclass[conference]{IEEEtran}
\IEEEoverridecommandlockouts
\usepackage{cite}
\usepackage{amsmath,amssymb,amsfonts}
\usepackage{algorithmic}
\usepackage{graphicx}
\usepackage{textcomp}
\usepackage{xcolor}
\usepackage{mathtools}
\usepackage{subcaption}
\usepackage[capitalize]{cleveref}

\usepackage{color}
\definecolor{gray}{RGB}{128,128,128}

\newcommand{\remove}[1]{} 

\makeatletter
\newcommand{\linebreakand}{%
  \end{@IEEEauthorhalign}
  \hfill\mbox{}\par
  \mbox{}\hfill\begin{@IEEEauthorhalign}
}
\makeatother

\usepackage{subfig}
\usepackage{booktabs}
\def\BibTeX{{\rm B\kern-.05em{\sc i\kern-.025em b}\kern-.08em
    T\kern-.1667em\lower.7ex\hbox{E}\kern-.125emX}}
\begin{document}

\title{Dynamic Operational Reserve Margin Assessment
from Risk-Constrained Unit Commitment States\\
\thanks{This manuscript has been authored by UT-Battelle, LLC, under contract DE-AC05-00OR22725 with the U.S. Department of Energy (DOE). The U.S. government retains and the publisher, by accepting the article for publication, acknowledges that the U.S. government retains a nonexclusive, paid-up, irrevocable, worldwide license to publish or reproduce the published form of this manuscript, or allow others to do so, for U.S. government purposes. DOE will provide public access to these results of federally sponsored research in accordance with the DOE Public Access Plan (https://www.energy.gov/doe-public-access-plan).}
}

\author{\IEEEauthorblockN{Pratishtha Shukla}
\IEEEauthorblockA{\textit{Computational Science \& Engineering}\\
\textit{Oak Ridge National Lab}\\
Oak Ridge, TN, USA \\
shuklap@ornl.gov}
\and
\IEEEauthorblockN{Shaked Regev}
\IEEEauthorblockA{\textit{Computational Science \& Engineering}\\
\textit{Oak Ridge National Lab}\\
Oak Ridge, TN, USA \\
regevs@ornl.gov}
\and
\IEEEauthorblockN{Evan J. R. Brody}
\IEEEauthorblockA{\textit{Computational Science \& Engineering} \\
\textit{Oak Ridge National Lab}\\
Oak Ridge, TN, USA \\
brodyej@ornl.gov}
\linebreakand
\IEEEauthorblockN{Charles Foltz}
\IEEEauthorblockA{\textit{Computer Science \& Mathematics} \\
\textit{Oak Ridge National Lab}\\
Oak Ridge, TN, USA \\
foltzcj1@ornl.gov}
\and
\IEEEauthorblockN{Teja Kuruganti}
\IEEEauthorblockA{\textit{Computational Science \& Engineering} \\
\textit{Oak Ridge National Lab}\\
Oak Ridge, TN, USA \\
kurugantipv@ornl.gov}
}

\maketitle

\begin{abstract}
We propose Dynamic Reserve Margin (DRM) as a time-varying operational adequacy metric derived from risk-constrained unit commitment (RCUC) states. DRM quantifies reserve adequacy using the additional generation capacity that committed generators can provide within a 5-minute response window relative to uncertainty and contingency reserve requirements. We introduce a complementary Reserve Risk Envelope (RRE) metric to quantify operational reserve headroom in directly interpretable MW terms. A low-margin duration metric is further developed to quantify the persistence of reserve stress over an operating horizon. We use an IEEE 14-bus example to illustrate these concepts, followed by large-scale RCUC case studies under multiple operating scenarios. Results demonstrate that reserve requirements and ramp-accessible reserve capability can vary substantially across operating conditions, and that commitment decisions adapt to maintain reserve adequacy under changing system conditions. The proposed DRM and RRE metrics provide an interpretable operational characterization of reserve adequacy, reveal reserve accessibility and stress persistence that are not captured by conventional reserve margin metrics.
\end{abstract}


\section{Introduction}
Reserve margin has traditionally been used as a planning-oriented adequacy metric to evaluate whether sufficient generation capacity exists to satisfy forecast demand and contingency requirements \cite{wood2013power,NERC}. In its conventional form, reserve margin is defined as the percentage of installed capacity exceeding expected peak demand \cite{NERC}. While this formulation provides a long-term indicator of resource sufficiency, it does not explicitly capture operational characteristics such as generator commitment state, ramping capability, uncertainty realization, or short-term reserve accessibility. Consequently, systems with apparently sufficient static reserve margins may nevertheless experience periods of operational reserve stress during rapidly changing operating conditions \cite{nycander2021security}.

As modern power systems experience increasing operational variability and uncertainty, the distinction between installed reserve capacity and operationally accessible reserve becomes more important \cite{nycander2021security}. In practice, reserve adequacy depends not only on how much capacity the system has, but also on how much additional generation the online generators can provide within operational response times \cite{park2022unit}. Under stressed operating conditions, generators may have limited ramping capability or insufficient headroom despite the presence of nominal reserve capacity \cite{parker2024managing}. Therefore, static reserve margin metrics may overestimate real-time operational reserve adequacy.

Risk-constrained unit commitment (RCUC) formulations \cite{regev2026oneperiodUC,kim2022reserve,nycander2021security,regev2026mechanicalUC,chen2022security} already capture many of these operational characteristics through commitment decisions, reserve constraints, ramping limits, and shortfall variables. While system operators such as PJM, MISO, and ERCOT monitor operational reserve availability in real time, these monitoring processes are generally implemented separately from the underlying unit commitment optimization. The resulting operational reserve feasibility information within RCUC therefore remains embedded implicitly in the optimization outputs and is not transformed into a consistent operational reserve adequacy metric derived directly from the commitment states. This motivates the need for a dynamic reserve characterization approach derived directly from RCUC operational states.

This paper proposes a Dynamic Reserve Margin (DRM) methodology that characterizes operational reserve adequacy using the additional generation capacity that committed generators can provide within a 5-minute response window relative to uncertainty and contingency reserve requirements. A complementary Reserve Risk Envelope (RRE) metric is introduced to quantify operational reserve headroom in directly interpretable MW terms, while a low-margin duration metric is used to characterize the persistence of reserve stress. The proposed methodology is first illustrated using an IEEE 14-bus example to demonstrate the distinction between planning-oriented and operational reserve adequacy. The methodology is then applied to large-scale RCUC \cite{regev2026mechanicalUC, regev2026oneperiodUC} case studies under multiple operating scenarios to evaluate operational reserve adequacy, reserve accessibility, and reserve requirement evolution under changing operating conditions.

\section{Dynamic Reserve Margin}
Conventional reserve margin provides a planning-oriented measure of resource adequacy by comparing installed generation capacity against forecast peak demand \cite{NERC}. While useful for long-term adequacy assessment, reserve margin does not directly quantify the amount of reserve that can be operationally accessed within the response times relevant to real-time system operation. Consequently, a distinction exists between installed reserve capacity and reserve that can be delivered by online generators over operational timescales.

The proposed DRM addresses this distinction by characterizing reserve adequacy using ramp-accessible reserve rather than installed capacity. Unlike static reserve margin, DRM depends on generator commitment decisions, dispatch levels, reserve allocations, ramp-rate limitations, and operating conditions \cite{krommydas2022flexibility}. The resulting metric therefore provides a time-varying characterization of operational reserve adequacy.

The conventional static reserve margin \cite{reimers2019impact} is defined as
\begin{equation}
RM^{\mathrm{static}}
=
\frac{
C^{\mathrm{installed}} - D^{\mathrm{peak}}
}{
D^{\mathrm{peak}}
},
\label{eq:static_rm}
\end{equation}
where $C^{\mathrm{installed}}$ denotes the total installed generation capacity and $D^{\mathrm{peak}}$ denotes the forecast peak demand. While this formulation provides a planning-oriented indicator of resource sufficiency, it does not account for operational feasibility constraints or short-term reserve accessibility.

In RCUC formulations, operational infeasibility under uncertainty is represented through reserve shortfall variables $s_t^{\omega} \geq 0$, where positive shortfall indicates that the committed fleet cannot satisfy reserve requirements  under scenario $\omega$ at interval $t$ \cite{regev2026mechanicalUC}. A system operating with acceptable risk maintains shortfall below a prescribed threshold, characterized by a risk measure $\rho(s_t^{\omega}) \leq \epsilon$, where $\rho(\cdot)$ may represent expected shortfall $\mathbb{E}_{\omega}[s_t^{\omega}]$, shortfall probability $\mathbb{P}(s_t^{\omega} > 0)$, or conditional value-at-risk $\text{CVaR}_{\alpha}(s_t^{\omega})$ \cite{geng2019chance, zhang2023unit}. However, RCUC outputs do not directly expose how much operational headroom exists above this threshold, that is, how far the system is from reserve deficiency at any given interval. This motivates a dynamic metric that quantifies ramp-accessible reserve headroom directly from RCUC operational states. 

The minimum reserve headroom required to satisfy a prescribed risk criterion may be defined as
\begin{equation}
    H_t^{\text{risk}} = \min_{\Delta_t \geq 0} 
    \left\{ \Delta_t : \rho\left(s_t^{\omega}(\Delta_t)\right) 
    \leq \epsilon \right\},
    \label{eq:hrisk}
\end{equation}
where $s_t^{\omega}(\Delta_t)$ denotes the reserve shortfall that would result after increasing the available ramp-accessible reserve by $\Delta_t$. This quantity characterizes the reserve headroom required to satisfy a prescribed risk criterion, but not the realized reserve accessibility provided by the committed generators. Note that the RCUC formulation of \cite{regev2026mechanicalUC} employs expected shortfall $\mathbb{E}_\omega[s_t^\omega]$ as the operational risk measure, and the proposed metrics are interpreted relative to this risk criterion throughout.

To quantify realized operational reserve adequacy, this work introduces a DRM formulation based on the amount of additional generation capacity that committed generators can provide within a 5-minute response window. Let the operational reserve requirement be defined as
\begin{equation}
R_t^{\mathrm{req}}
=
\Delta_t^{\mathrm{unc}}
+
\Delta_t^{\mathrm{cont}},
\label{eq:reserve_requirement}
\end{equation}
where $\Delta_t^{\mathrm{unc}}$ represents reserve requirements associated with forecast uncertainty and operational variability, and $\Delta_t^{\mathrm{cont}}$ represents contingency reserve requirements associated with generator outages or system disturbances.

The total ramp-accessible reserve available within a 5-minute response window is then defined as
\begin{equation}
R_t^{5\mathrm{min}}
=
\sum_{i \in \mathcal{G}}
u_{i,t}
\min
\left(
P_i^{\max} - P_{i,t},
5r_i^{\mathrm{up}}
\right),
\label{eq:ramp_accessible_reserve}
\end{equation}
where $\mathcal{G}$ denotes the generator set, $u_{i,t}$ is an indicator variable denoting the commitment state of generator $i$ at time $t$, $P_i^{\max}$ denotes the maximum generation capacity, $P_{i,t}$ denotes the current dispatch level, and $r_i^{\mathrm{up}}$ denotes the generator ramp-up capability in MW/min. The quantity $R_t^{5\mathrm{min}}$ therefore represents the total additional generation that the online generators can provide within a 5-minute operational response window. While reserve products and response requirements vary by market and timescale, the 5-minute window represents a standard operational benchmark for short-term reserve accessibility assessment \cite{NERC}. The proposed framework generalizes directly to other response windows by adjusting the scalar multiplier 
in \cref{eq:ramp_accessible_reserve}.

Using the ramp-accessible reserve and operational reserve requirement, DRM is defined as
\begin{equation}
DRM_t
=
\frac{
R_t^{5\mathrm{min}}
-
R_t^{\mathrm{req}}
}{
D_t
},
\label{eq:drm}
\end{equation}
where $D_t$ denotes the system demand at operating interval $t$. Unlike static reserve margin, the proposed DRM is time-varying and depends explicitly on commitment feasibility, generator headroom, ramping capability, and uncertainty conditions. Positive DRM values indicate sufficient operational reserve accessibility, while negative DRM values indicate reserve-deficient operating conditions in which the online generators cannot provide sufficient additional reserve within the operational response window.

To provide a directly interpretable operational reserve adequacy signal, this work further defines the RRE as
\begin{equation}
RRE_t
=
R_t^{5\mathrm{min}}
-
R_t^{\mathrm{req}}.
\label{eq:rre}
\end{equation}
The RRE quantifies the remaining operational reserve headroom in MW terms after accounting for uncertainty and contingency reserve requirements. Positive RRE values indicate adequate operational reserve conditions, near-zero values indicate stressed operating conditions, and negative values indicate reserve-deficient operating intervals. This interpretation may be formalized as
\begin{equation}
\text{Operating State}_t=
\begin{cases}
\text{Adequate}, & RRE_t > \delta, \\
\text{Stressed}, & |RRE_t| \leq \delta, \\
\text{Reserve Deficient}, & RRE_t < -\delta,
\end{cases}
\label{eq:rre_state}
\end{equation}
where $\delta$ denotes a small reserve tolerance threshold that accounts for operational uncertainty and distinguishes reserve-balanced operating conditions from clearly adequate or reserve-deficient states. In practice, $\delta$ may be selected as a fixed MW value or as a small percentage of system demand.

The proposed methodology additionally enables decomposing operational reserve adequacy into reserve requirements and ramp-accessible reserve capability. This decomposition provides insight into whether changes in reserve adequacy arise from increasing reserve requirements, changes in commitment patterns, or changes in operational reserve accessibility.

Finally, to characterize reserve adequacy under uncertainty, the DRM and RRE are evaluated across multiple operating scenarios. For scenario $\omega \in \Omega$, the scenario-conditioned DRM and RRE are defined as
\begin{align}
DRM_t^{\omega}
&=
\frac{
R_t^{5\mathrm{min},\omega}
-
R_t^{\mathrm{req},\omega}
}{
D_t^{\omega}
},
\\
RRE_t^{\omega}
&=
R_t^{5\mathrm{min},\omega}
-
R_t^{\mathrm{req},\omega}.
\end{align}
These quantities enable construction of reserve adequacy envelopes and characterization of reserve stress trajectories.

DRM may further be used to quantify the persistence of low reserve adequacy conditions over an operating horizon. Let $\theta$ denote a minimum acceptable operational reserve margin threshold and let $\mathcal{T}$ denote the set of operating intervals. The low-margin duration is defined as
\begin{equation}
SD^\omega
=
\frac{1}{|\mathcal{T}|}
\sum_{t \in \mathcal{T}}
\mathbf{1}
\left(
DRM_t^\omega \leq \theta
\right),
\label{eq:stress_duration}
\end{equation}
where $\mathbf{1}(\cdot)$ is the indicator function. The resulting quantity represents the fraction of the operating horizon during which the DRM remains below a prescribed operational adequacy threshold. Larger values of $SD^\omega$ indicate reduced reserve accessibility and increased persistence of operational reserve stress. In this work, $\theta=10\%$ is used as a representative
threshold for identifying intervals with limited
operational reserve headroom.


\section{Methodology and Implementation}
\subsection{DRM Evaluation Methodology}
The proposed DRM methodology operates as a post-processing operational adequacy assessment derived from RCUC states. The underlying RCUC formulation determines generator commitment decisions $u_{i,t}$, dispatch levels $P_{i,t}$, reserve allocations $R_{i,t}$, and reserve shortfall variables $s_t^{\omega}$ for each operating interval and uncertainty scenario.  Since the objective of this work is not to develop a new unit commitment formulation, the details of the underlying optimization are omitted and the reader is referred to \cite{regev2026mechanicalUC}. Using these outputs, the proposed DRM methodology evaluates the amount of reserve that can be operationally accessed within a specified response interval.



For each operating interval, the total operational reserve requirement is computed as the sum of uncertainty and contingency reserve requirements. The amount of reserve that can be accessed within a 5-minute response interval is subsequently evaluated using committed generator states, available generation headroom, and generator ramp-rate capabilities. For each committed generator, the accessible reserve is limited by both the unused generation headroom and the amount of generation that can be delivered within the 5-minute response horizon.

\begin{figure}[t]
    \centering
    \includegraphics[width=0.85\columnwidth]{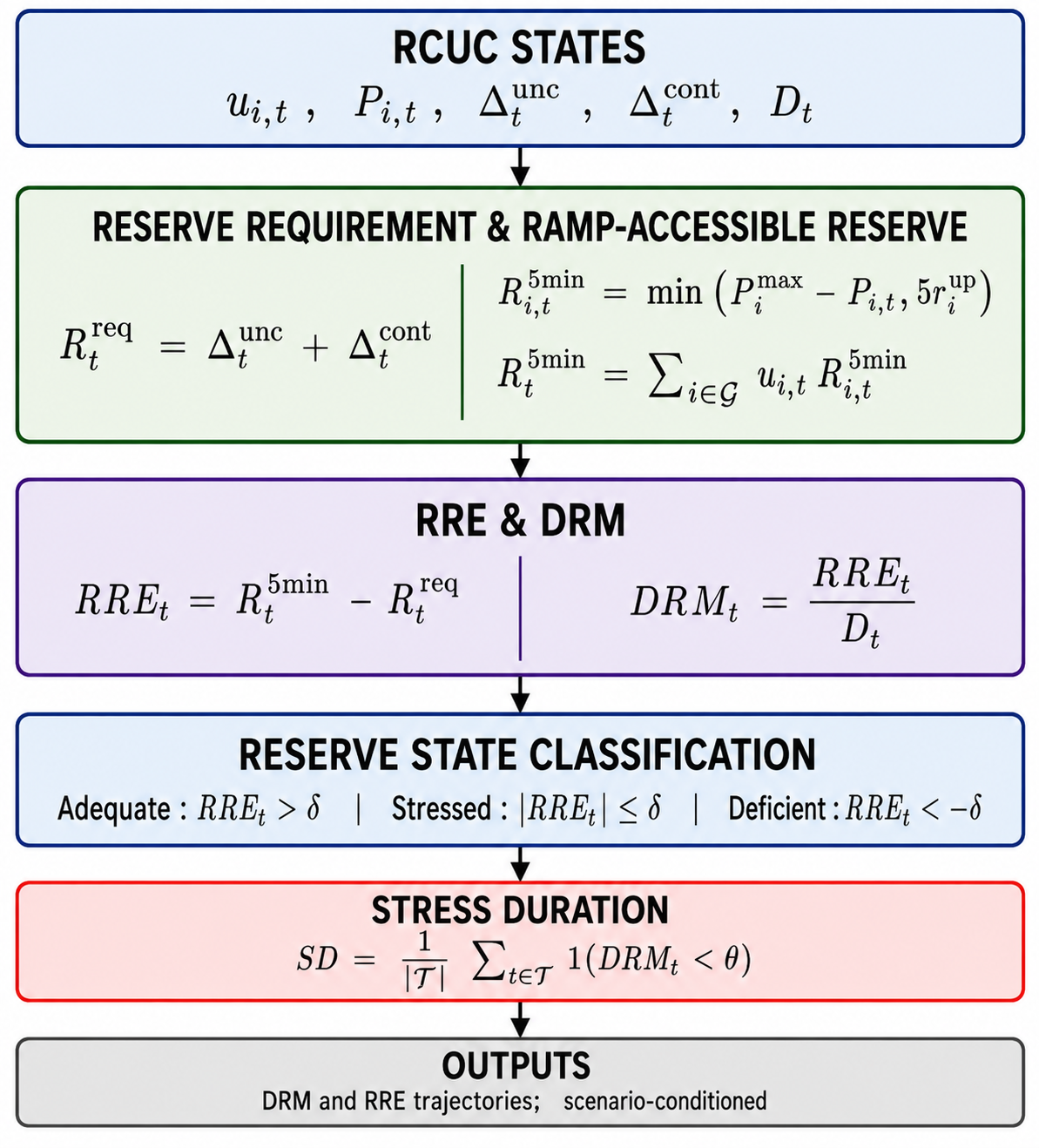}
    \caption{Overview of the DRM evaluation process.}
    \label{fig:drm_workflow}
\end{figure}

\cref{fig:drm_workflow} shows the DRM evaluation workflow. RCUC operational states are processed to determine reserve requirements and ramp-accessible reserve capability for each operating interval. The total ramp-accessible reserve is obtained by aggregating the accessible reserve contribution of all committed generators. These quantities are subsequently used to evaluate DRM, RRE, and reserve stress metrics, yielding time-varying trajectories. The resulting trajectories provide a time-varying characterization of operational reserve adequacy and interactions between reserve requirements and ramp-accessible reserve over the operating horizon. 

\subsection{Study Systems and Scenarios}
\subsubsection{Illustrative Example} An IEEE 14-bus system \cite{IEEE14} is first used as an illustrative proof-of-concept example to demonstrate the proposed DRM and RRE concepts under controlled operating conditions. The IEEE 14-bus RAW file is parsed to extract system topology, load information, and generator locations. A 24-hour operating profile is subsequently generated using normalized load multipliers to create low-load, ramping, and peak-demand operating periods. Simplified generator capacities and ramp-rate assumptions are assigned to emulate representative operational reserve behavior. Additional stress conditions are introduced through uncertainty reserve requirements, contingency reserve requirements, and generator availability reductions. The objective of this illustrative example is not to perform a fully realistic operational study, but rather to visualize the distinction between planning-oriented reserve adequacy and operational reserve adequacy under stressed operating conditions.

\subsubsection{Large-Scale Case Study}
The proposed methodology is further evaluated using RCUC simulations on a large-scale single-node system with generator data derived from PJM Data Miner \cite{PJMDataMiner}, following the setup in \cite{regev2026mechanicalUC}. The system includes 462 generators and a 24-hour operating horizon represented at 5-minute resolution. Generator commitment status, dispatch level, and reachable generation within a 5-minute response interval are obtained from the RCUC outputs for each scenario. 
The resulting operational states are subsequently processed using the proposed DRM methodology to evaluate operational reserve adequacy, reserve accessibility, reserve requirement evolution, and reserve stress persistence across scenarios.

\section{Results and Discussion}
This section evaluates the DRM and RRE metrics using the IEEE 14-bus example and large-scale grid RCUC simulations. 

\subsection{IEEE 14-Bus Illustrative Example}
The DRM methodology is evaluated on an IEEE 14-bus system \cite{IEEE14} under time-varying demand, uncertainty reserve requirements, contingency reserve requirements, and representative outage conditions over a 24-hour operating horizon. Simplified generator commitment and dispatch assumptions are used to emulate operational reserve behavior under stressed conditions. The objective of this illustrative example is to demonstrate the distinction between planning-oriented reserve adequacy and operational reserve accessibility within a 5-minute response window.

For comparison purposes, a time-varying conventional reserve margin estimate is additionally computed using the instantaneous system demand,
$
RM_t^{\mathrm{conv}}
=
(
C^{\mathrm{installed}} - D_t
)/
D_t.
$
While this quantity does not represent the traditional planning reserve margin defined relative to  peak demand in \cref{eq:static_rm}, it provides a useful reference for comparing installed reserve capacity against the proposed operational DRM trajectories over the operating horizon.

\begin{figure}[!t]
\centering
\begin{minipage}{0.48\columnwidth}
    \centering
    \includegraphics[width=\linewidth]{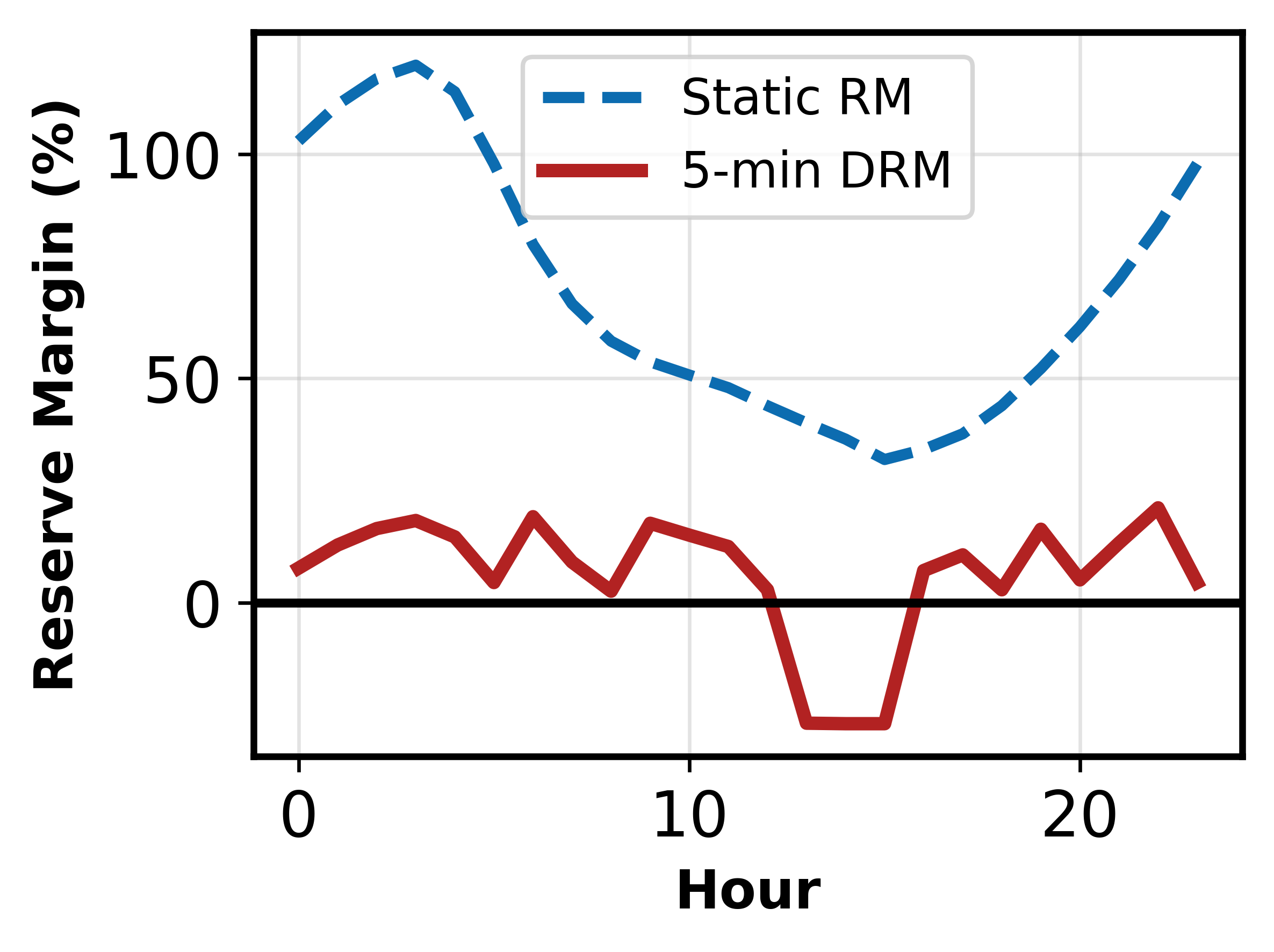}
    \subcaption{Static vs. dynamic DRM}
    \label{fig:static_dynamic_rm}
\end{minipage}
\hfill
\begin{minipage}{0.48\columnwidth}
    \centering
    \includegraphics[width=\linewidth]{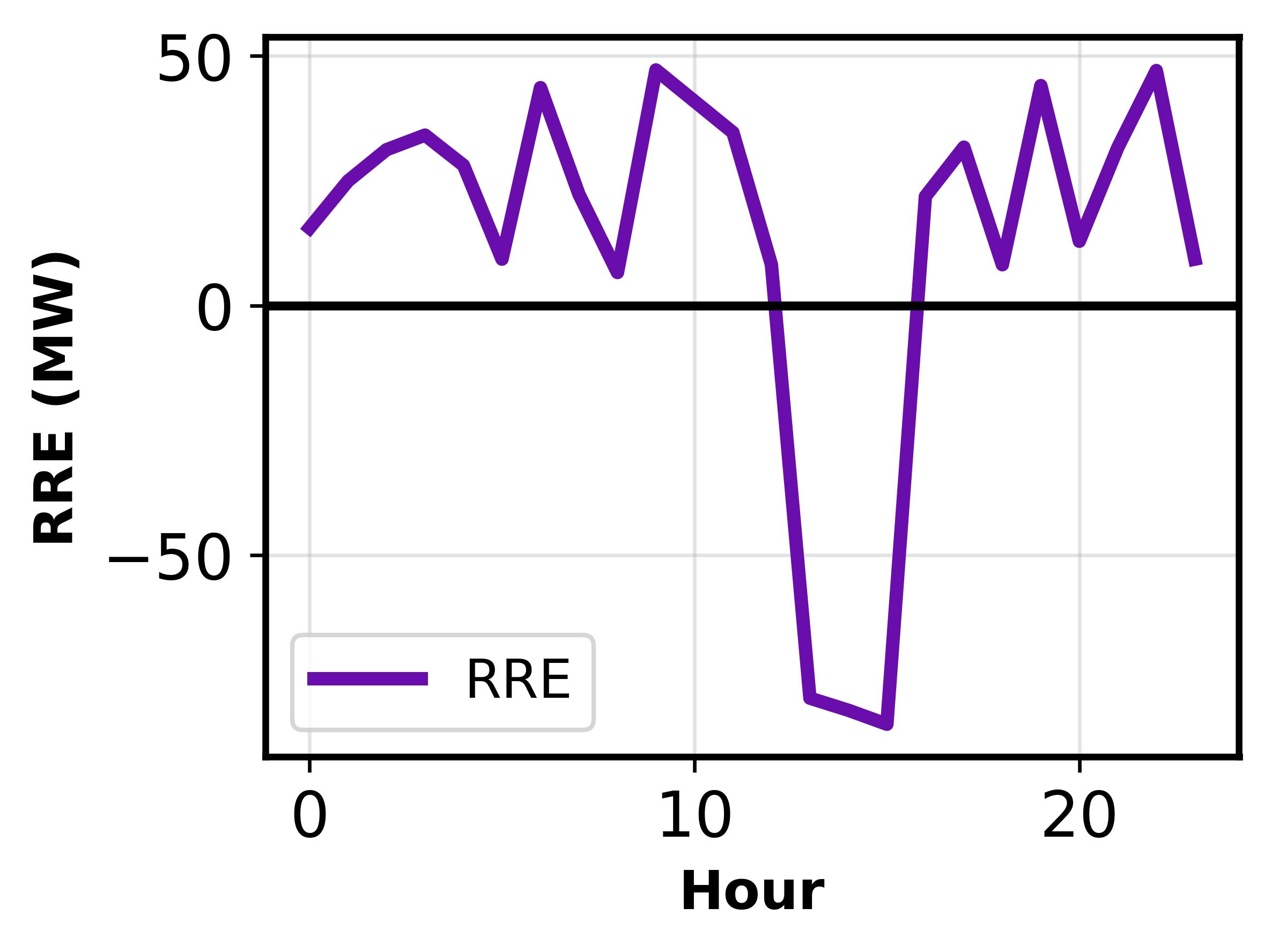}
    \subcaption{Reserve Risk Envelope}
    \label{fig:rre}
\end{minipage}
\caption{Static RM, DRM, and RRE trajectories for the IEEE 14-bus example.}
\label{fig:drm_rre}
\end{figure}

\cref{fig:drm_rre} compares the conventional static reserve margin against the proposed 5-minute DRM and corresponding RRE trajectory. The static reserve margin remains positive throughout the operating horizon due to the large installed generation capacity relative to system demand. However, the proposed DRM reveals periods of operational reserve stress despite the apparently sufficient planning reserve margin. This illustrates that installed reserve capacity alone does not guarantee operational reserve accessibility within short response horizons. In particular, the stressed operating interval around hours 13--15 coincides with the peak-load period of the operating profile. During this interval, the DRM becomes negative, indicating that the online generators cannot provide sufficient additional reserve within the 5-minute response window to satisfy uncertainty and contingency reserve requirements. The RRE trajectory  in \cref{fig:drm_rre} provides a directly interpretable operational reserve adequacy signal in MWs. Positive RRE values indicate sufficient ramp-accessible reserve headroom, while negative RRE values indicate reserve-deficient operating conditions. The negative RRE interval during the stressed operating period illustrates the loss of operational reserve accessibility despite the presence of sufficient installed generation capacity from a planning perspective.

\begin{figure}[!t]
\centering

\begin{minipage}{0.48\columnwidth}
    \centering
    \includegraphics[width=\linewidth]{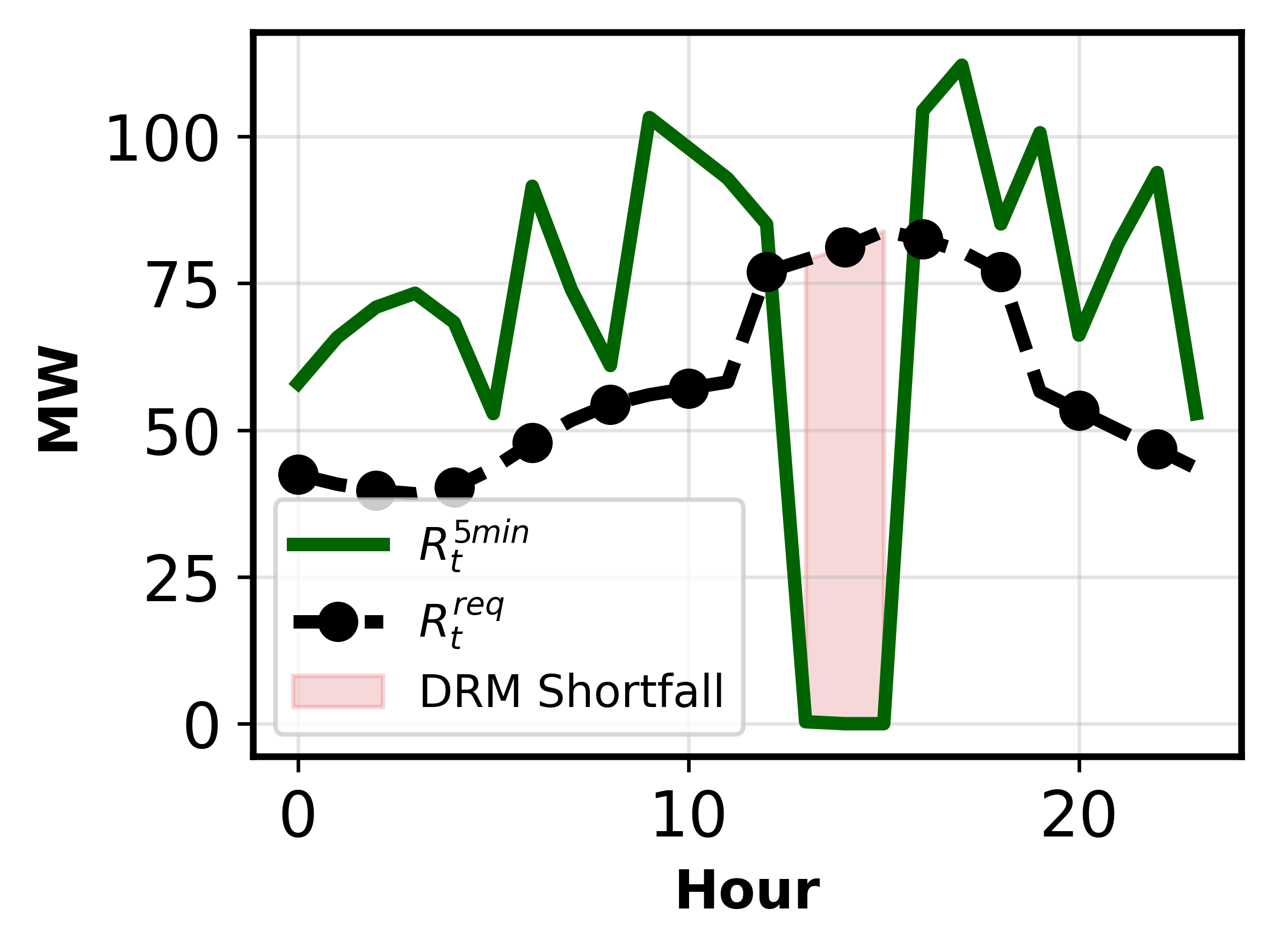}
    \subcaption{Risk-Adjusted Demand}
    \label{fig:risk_adjusted_demand}
\end{minipage}
\hfill
\begin{minipage}{0.48\columnwidth}
    \centering
    \includegraphics[width=\linewidth]{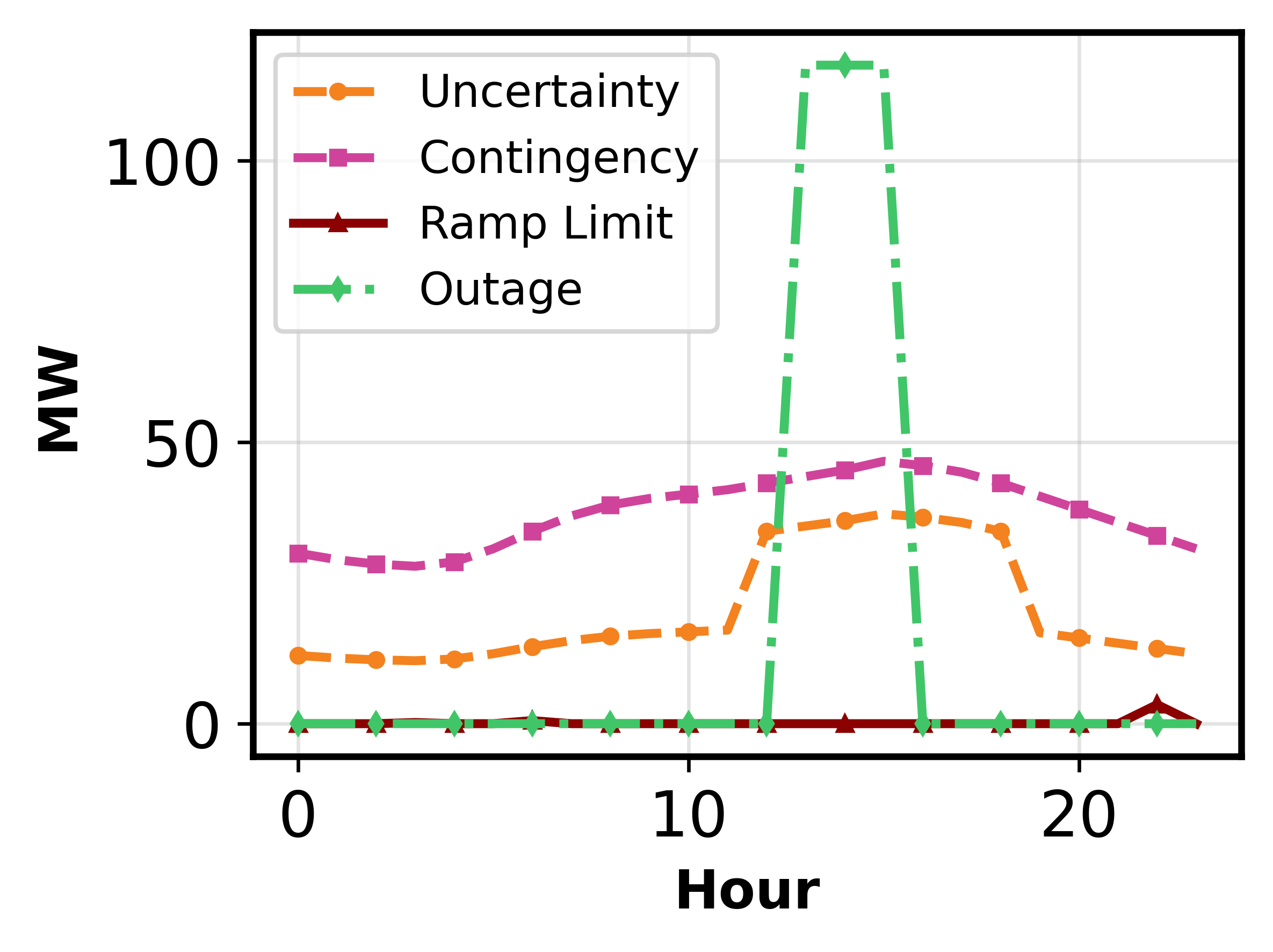}
    \subcaption{DRM Driver Decomposition}
    \label{fig:driver_decomp}
\end{minipage}

\caption{Ramp-accessible reserve, reserve requirement, and DRM driver decomposition for the IEEE 14-bus example.}
\label{fig:risk_driver}
\end{figure}

\cref{fig:risk_driver} illustrates the relationship between ramp-accessible reserve capability and operational reserve requirements. The reserve requirement increases during high-stress operating periods while the available 5-minute ramp-accessible reserve decreases due to outage conditions and reduced generator flexibility. The shaded interval highlights periods in which ramp-accessible reserve falls below the operational reserve requirement, resulting in both DRM $<0$, and RRE $<0$.

The driver decomposition results in \cref{fig:risk_driver} illustrate the operational mechanisms contributing to DRM degradation. There, outage conditions and reduced ramp-accessible reserve are the primary drivers of reserve stress, while uncertainty and contingency reserve requirements increase reserve demand during higher-load periods. These results demonstrate that operational reserve adequacy depends not only on installed reserve capacity, but also on generator availability, ramping capability, and operational reserve accessibility.


\subsection{Case Study}
The proposed DRM methodology was evaluated using risk-constrained unit commitment (RCUC) simulations on a large-scale single-node system with generator characteristics derived from PJM Data Miner data \cite{PJMDataMiner}. The study considers five operating conditions representing progressively increasing system stress:
(1) Base operation, (2) Generator outage, (3) High load, (4) High uncertainty, and (5) Combined stress.

For each scenario, available ramp-accessible reserve was computed from generator commitment, dispatch, and 5-minute ramping capability. DRM and RRE values were then evaluated at each operating interval.

\begin{figure} [h]
\centering
\includegraphics[width=\columnwidth]{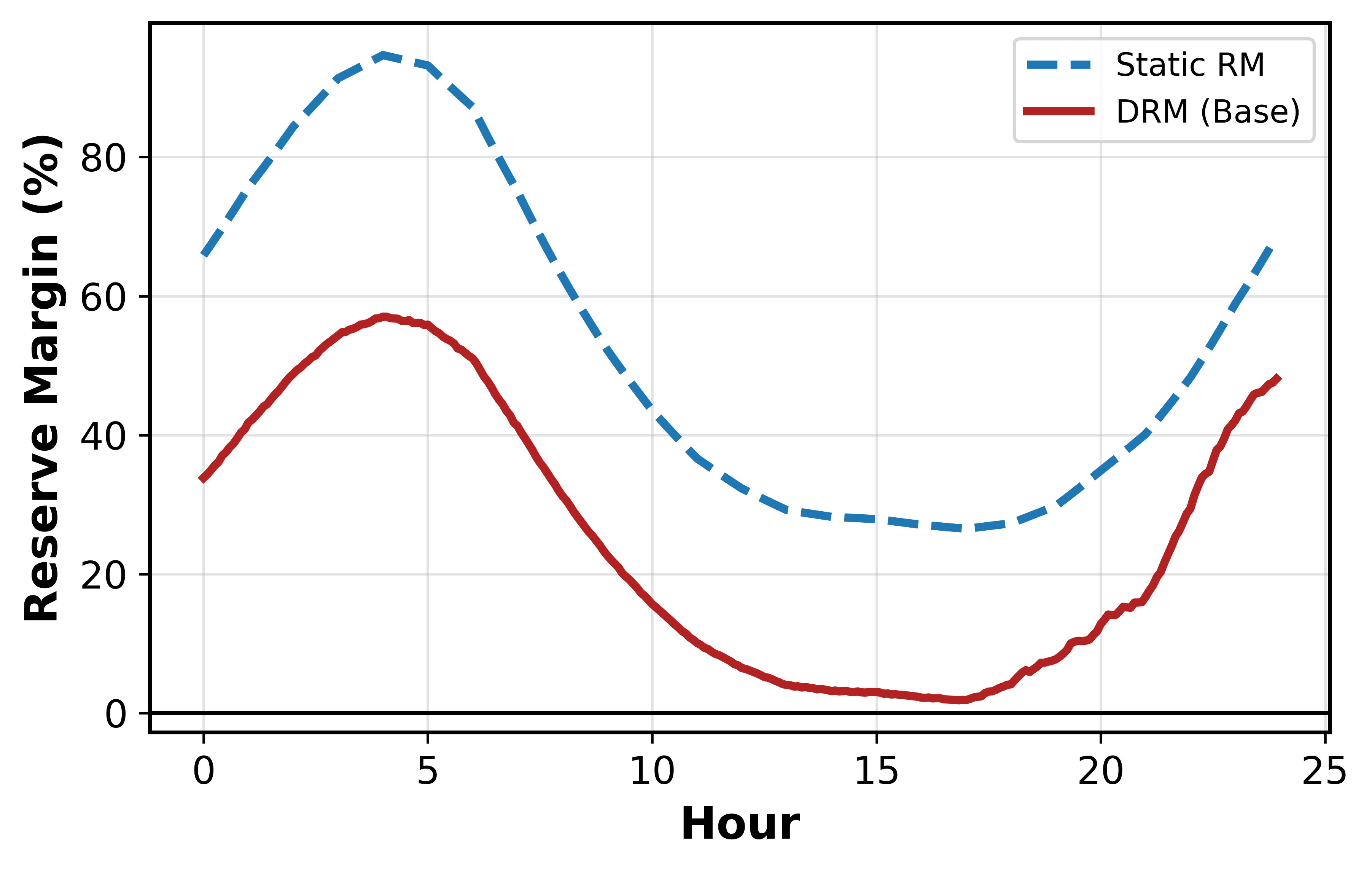}
\caption{Comparison of static reserve margin and DRM for the base scenario.}
\label{fig:static_vs_dynamic_rm}
\end{figure}
\cref{fig:static_vs_dynamic_rm} compares the conventional static reserve margin with the proposed DRM for the base scenario. The static reserve margin assumes that all installed capacity is immediately available and therefore remains substantially larger than the operationally achievable reserve margin throughout the day. In contrast, DRM reflects commitment status, dispatch levels, and ramp-rate limitations, producing a more conservative but operationally realistic assessment of reserve adequacy. The largest discrepancies occur during periods of elevated demand when available ramping capability becomes constrained.

\begin{figure}[t]
\centering
\includegraphics[width=\columnwidth]{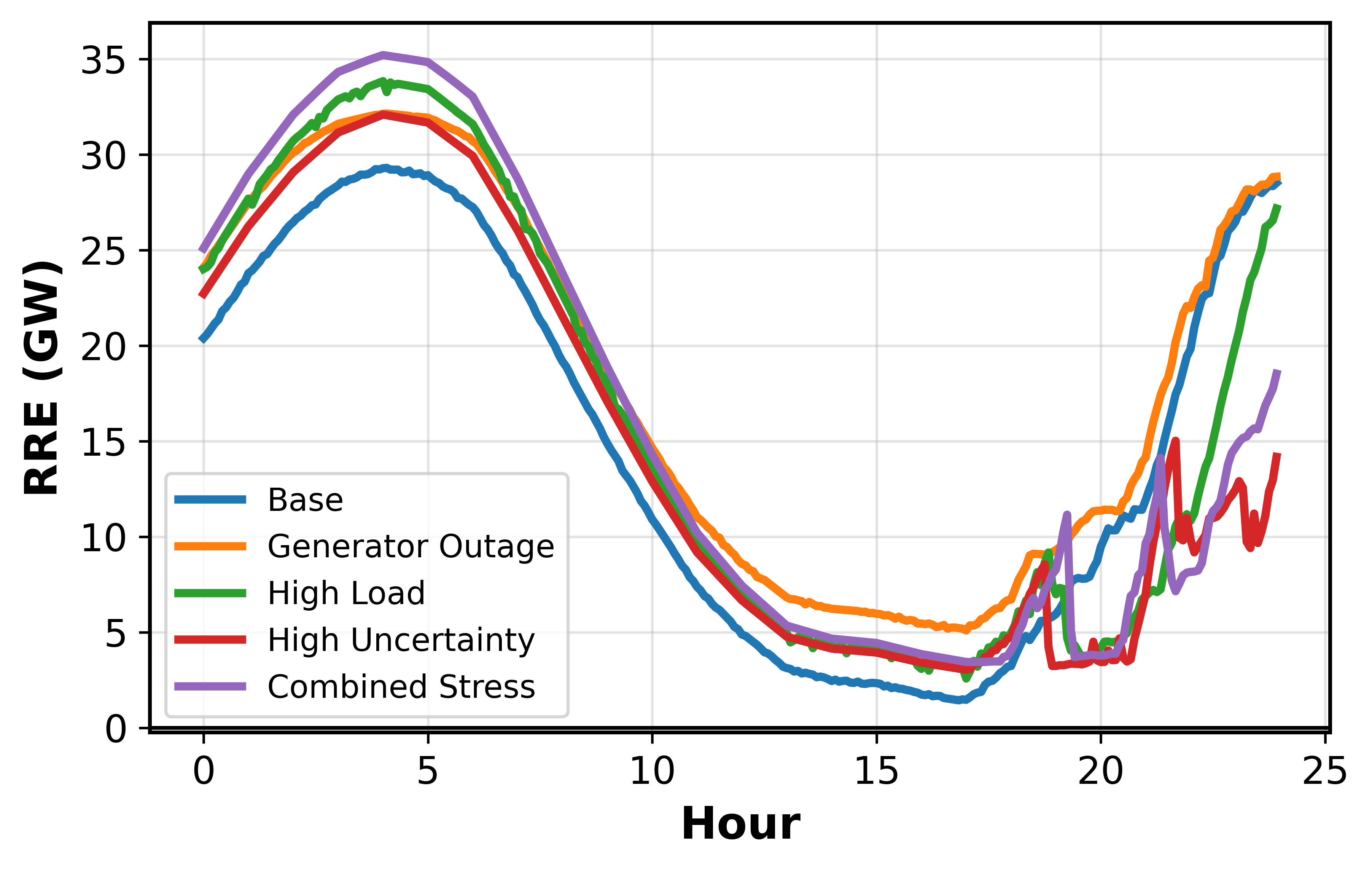}
\caption{RRE trajectories across operating scenarios.}
\label{fig:rre_scenarios}
\end{figure}
\cref{fig:rre_scenarios} presents the resulting RRE trajectories across all operating scenarios. Positive RRE values indicate that reachable reserves exceed reserve requirements, while values approaching zero indicate increasing reserve stress. Although all scenarios maintain positive RRE values throughout the operating horizon, noticeable differences emerge in the magnitude and temporal evolution of reserve headroom. The high-load, high-uncertainty, and combined-stress scenarios experience reduced reserve headroom during peak operating periods, reflecting the increased reserve requirements imposed by loading and uncertainty conditions.

\begin{figure}[t]
\centering
\includegraphics[width=\columnwidth]{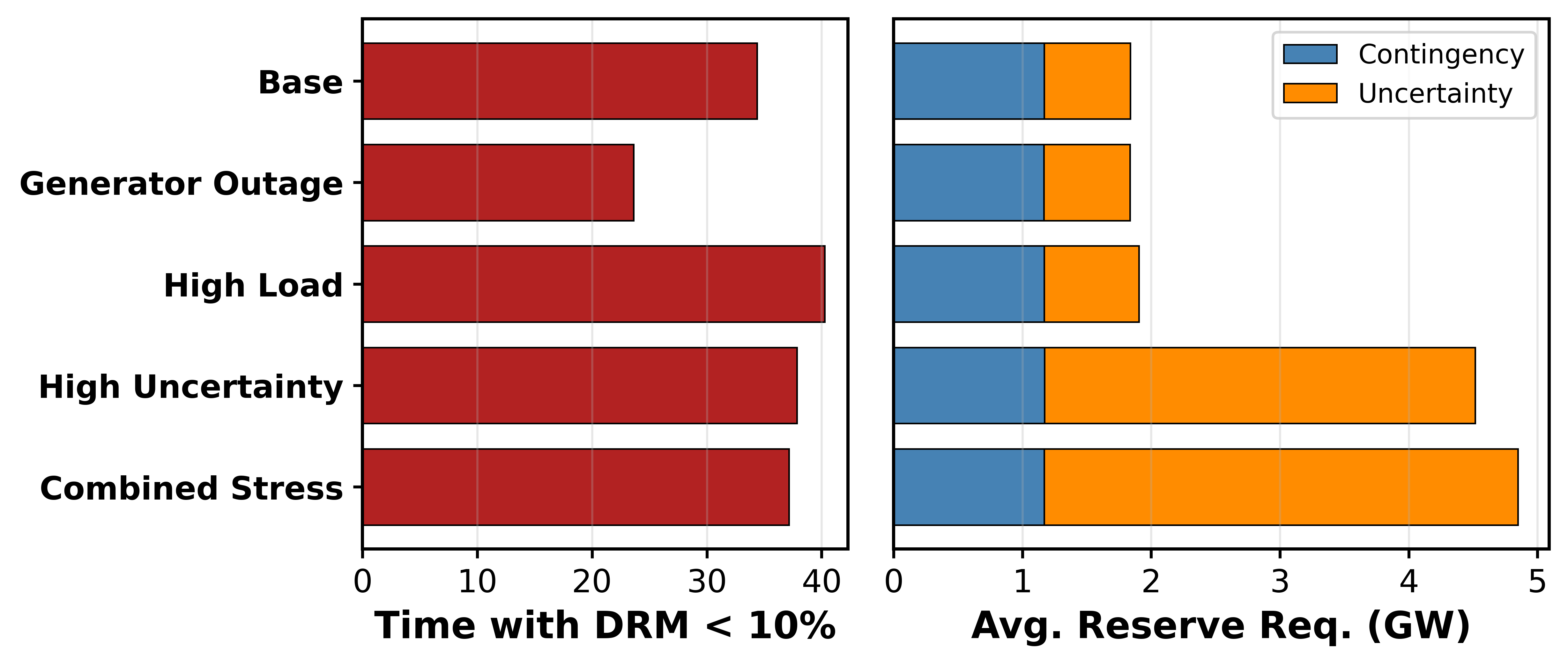}
\caption{Scenario-based reserve adequacy assessment. (a) Fraction of the operating horizon with DRM below the low-margin threshold. (b) Average reserve requirement decomposition.}
\label{fig:scenario_assessment}
\end{figure}

To quantify the persistence of reserve stress, a low-margin duration metric was computed using a DRM threshold of 10\%. \cref{fig:scenario_assessment}(a) summarizes the fraction of the operating horizon during which the system operated with DRM below this threshold. 
The high-load, high-uncertainty, and combined-stress scenarios exhibit the largest low-margin durations. These also experience reduced reserve headroom during the afternoon peak-load period, reflecting the increased reserve requirements imposed by loading and uncertainty conditions.
In contrast, the generator-outage scenario experiences the shortest low-margin duration. This indicates that loading and uncertainty conditions contribute more strongly to persistent reductions in reserve margin than the outage scenario considered in this study.

To further investigate the underlying causes of reserve stress, \cref{fig:scenario_assessment}(b) decomposes the average reserve requirement into contingency and uncertainty components for each operating scenario. The contingency requirement remains relatively constant because it is determined by the largest online generator and therefore changes little across scenarios. In contrast, the uncertainty component increases substantially under the high-uncertainty and combined-stress scenarios, becoming the dominant contributor to reserve requirement growth. The increase in uncertainty reserve requirements is accompanied by a corresponding increase in ramp-accessible reserve through RCUC commitment adaptation. As a result, average DRM values remain relatively stable across scenarios despite substantial changes in reserve requirements.

\cref{tab:scenario_summary} summarizes key reserve adequacy metrics across all operating scenarios. The average DRM remains relatively consistent across scenarios, ranging from approximately 25\% to 30\%, indicating that the tested operating conditions maintain adequate reserve accessibility over most of the operating horizon. However, the low-margin duration metric reveals meaningful differences in reserve stress persistence. The high-load scenario exhibits the largest low-margin duration, followed closely by the high-uncertainty and combined-stress scenarios, indicating prolonged periods during which operational reserve margins are reduced despite remaining positive. In contrast, the generator-outage scenario exhibits the shortest low-margin duration and the highest average DRM. This behavior arises from the RCUC commitment response, which commits additional units to satisfy power balance
following generator removal. The resulting commitment pattern increases aggregate ramp-accessible reserve and illustrates that DRM reflects operational reserve accessibility rather than installed capacity alone.

Although average DRM values remain relatively similar across
scenarios, the underlying reserve requirements and accessible
reserve levels vary substantially. For example, average reserve requirements increase from approximately 1.8 GW in the base case to more than 4.5 GW in the high-uncertainty and combined-stress scenarios. However, the corresponding RCUC commitment decisions increase ramp-accessible reserve from approximately 16.8 GW to 19.3-21.2 GW, thereby maintaining  positive reserve margins across all scenarios. This behavior demonstrates the ability of the RCUC formulation to adapt commitment decisions in response to changing operating conditions while maintaining operational reserve adequacy.

Notably, all scenarios maintain positive minimum DRM and RRE values throughout the operating horizon. Consistent with the risk-constrained interpretation of \eqref{eq:hrisk}, reserve shortfall variables $s_t^{\omega}$ remained zero across all scenarios, indicating that reserve requirements were satisfied and the system operated below the prescribed risk threshold $\epsilon$. 
The scenarios considered in this study therefore illustrate how DRM distinguishes varying degrees of operational reserve adequacy within feasible RCUC operating conditions.

\begin{table}[t]
\caption{Scenario comparison using DRM-based metrics.}
\label{tab:scenario_summary}
\centering
\footnotesize
\begin{tabular}{lccc}
\toprule
Scenario &
Mean DRM &
Min DRM &
Low-Margin \\
&
(\%) &
(\%) &
(\%) \\
\midrule
Base             & 25.19 & 1.82 & 34.38 \\
Gen. Outage & 30.07 & 6.45 & 23.61 \\
High Load        & 24.89 & 2.97 & 40.28 \\
High Uncertainty & 25.09 & 3.84 & 37.85 \\
Combined Stress  & 25.27 & 3.96 & 37.15 \\
\bottomrule
\end{tabular}
\end{table}

\section{Conclusion}
This paper proposes a Dynamic Reserve Margin (DRM) methodology for operational reserve adequacy assessment using Risk-Constrained Unit Commitment (RCUC) states. DRM quantifies reserve adequacy using ramp-accessible generation capability available within a 5-minute response window relative to operational reserve requirements. A complementary RRE metric was introduced to provide an interpretable measure of operational reserve headroom. The methodology was demonstrated using an IEEE 14-bus example and a large-scale PJM-based RCUC case study. Results showed that reserve requirements and accessible reserve capability can vary substantially across operating conditions, while RCUC commitment decisions adapt to maintain operational reserve adequacy. The proposed DRM methodology provides a more operationally meaningful characterization of reserve adequacy than conventional reserve margin metrics by explicitly accounting for commitment status, reserve accessibility, and response-time constraints. Future work will extend the framework to network-constrained unit commitment models, probabilistic operational reliability assessment, and more extreme large-scale stress scenarios that explicitly enter reserve-deficient operating regimes.

\bibliographystyle{IEEEtran}
\bibliography{main}

\end{document}